\documentclass[
]{ceurart}

\usepackage{listings}
\begin{document}

\copyrightyear{2025}
\copyrightclause{Copyright for this paper by its authors.
  Use permitted under Creative Commons License Attribution 4.0
  International (CC BY 4.0).}

\conference{First International Workshop on Advancing AI Literacy with Learning Analytics (AI-LIT) @ LAK26, April 2026}

\title{Tracing GenAI Literacy: Uncovering Student-AI Interaction Patterns in Academic Writing through Epistemic Network Analysis}

\author[1]{Angxuan Chen}[]
\address[1]{Department of Educational Technology, Graduate School of Education, Peking University, Beijing, China}
\author[1]{Jiyou Jia}[%
email=jjy@pku.edu.cn,
]
\cormark[1]

\cortext[1]{Corresponding author.}
\begin{abstract}
  As Generative AI (GenAI) becomes integral to education, fostering
  GenAI literacy is critical. However, current assessments largely rely
  on self-reported scales, lacking insights into how literacy manifests
  in actual learning processes. This study leverages Learning Analytics
  (LA) to bridge this gap. We collected interaction logs from 162
  university students engaged in a GenAI-assisted abstract writing
  task. Using Epistemic Network Analysis (ENA), we modeled and compared
  the questioning strategies of students with varying GenAI literacy
  levels. Preliminary results reveal distinct interaction signatures:
  high-literacy students engage in iterative refinement and strategic
  questioning, while low-literacy students rely on direct generation
  commands. This work contributes to the workshop by demonstrating how
  process data can characterize GenAI literacy, paving the way for
  data-driven literacy assessment and real-time interventions.
\end{abstract}

\begin{keywords}
  GenAI literacy \sep
  Learning Analytics \sep
  Epistemic Network Analysis \sep
  Interaction Patterns \sep
  Process Data
\end{keywords}

\maketitle

\section{Introduction}

The integration of Generative AI (GenAI) into educational settings has
shifted the focus from mere access to ``GenAI Literacy''---a
multidimensional competence encompassing the understanding, application,
evaluation, and ethical use of AI \cite{Chiu2024}. As GenAI tools like
ChatGPT and DeepSeek become ubiquitous, they offer the potential to
augment human cognition, but only if learners possess the requisite
literacy to navigate them effectively.

However, a significant gap exists in how GenAI literacy is currently
measured. Existing research predominantly treats literacy as a static
trait assessed via self-reported questionnaires (e.g.,
\cite{Almatrafi2024}). While useful, these instruments miss the
``process'' dimension: they cannot reveal how literacy manifests during
actual human-AI collaboration. Does a student with high self-reported
literacy actually prompt more effectively? Do low-literacy students
struggle with hallucination management in real-time?

To address this challenge, we propose a Learning Analytics (LA) approach
to characterize literacy through behavioral evidence. By analyzing
interaction logs via Epistemic Network Analysis (ENA), this study aims
to identify specific behavioral markers that distinguish high from low
GenAI literacy users. This research moves beyond self-reported confidence
to investigate how students actually structure their interaction with
GenAI during complex academic tasks, providing a foundation for objective
literacy assessment.

\section{Methodology}

\subsection{Context and Participants}

A total of 162 university students ($M_{\text{age}} = 20.1$)
participated in this study. Prior to the task, students completed a
validated GenAI Literacy Test \cite{Jin2025} assessing technical
understanding, interaction skills, and ethical awareness. Based on the
median score, participants were classified into a High Literacy Group
(HG, $n=89$) and a Low Literacy Group (LG, $n=73$).

The learning task required students to write an academic abstract for a
research paper using a custom-built platform integrated with the
DeepSeek Large Language Model \cite{DeepSeek2025}. To encourage
meaningful interaction, the platform imposed a 30-word limit per prompt
and disabled copy-pasting, forcing students to synthesize AI outputs
rather than passively adopting them.

\subsection{Data Collection and Coding Framework}

The primary data source was the full interaction log (timestamped prompts
and AI responses). We adapted a coding framework from Liu et al.
\cite{Liu2024} to categorize student prompts into specific communicative
acts. The framework distinguishes between Command Behaviors and
Questioning Behaviors:

\begin{table}
  \caption{Coding Framework}
  \label{tab:coding}
  \begin{tabular}{p{3.5cm}p{4.5cm}p{4.5cm}}
    \toprule
    Type & Description & Example \\
    \midrule
    Abstract Generation Command (AGC) & Direct requests to generate text & Write an abstract for me \\
    AI Improvement Command (AIC) & Requests to refine or polish text & Make this sentence more academic \\
    Meta-Command (MC) & Inquiries about how to prompt or use the tool & How should I summarize this \\
    Clarification Question (CQ) & Seeking explanation of concepts & What does `epistemic' mean here? \\
    Factual Question (FQ) & Retrieving specific facts from the text & What was the total sample size in the 2019 survey? \\
    Challenging Question (CHQ) & Questioning the AI's logic or accuracy & Are you sure about that percentage? \\
    \bottomrule
  \end{tabular}
\end{table}

Two researchers independently coded a subset of the data, achieving
substantial inter-rater reliability (Cohen's $\kappa = 0.75$).

\subsection{Epistemic Network Analysis (ENA)}

To capture the structure of interactions, we utilized Epistemic Network
Analysis (ENA) \cite{Shaffer2016}. Unlike frequency counts, ENA models
the co-occurrence of codes within a moving temporal window, visualizing
how different behaviors (e.g., asking for improvement + asking for
clarification) are linked in the students' cognitive process. This allows
for a structural comparison of interaction strategies between the HG and
LG groups.

\section{Preliminary Findings}

\subsection{Interaction Patterns via ENA}

The ENA models revealed structurally distinct interaction patterns
between the two groups. The Low Literacy (LG) network is dominated by
strong connections between Abstract Generation Commands (AGC) and
Factual Questions (FQ) or Challenging Questions (CHQ). This pattern
indicates a ``transactional'' approach. LG students tended to issue
broad commands for content generation (``Write this'') and paired them
with basic fact retrieval. The link to CHQ often appeared as reactions to
confusion or hallucinations rather than strategic critique. This
structure suggests a reliance on GenAI as a search engine or a black-box
generator rather than a collaborative agent.

\begin{figure}
  \centering
  \includegraphics[width=\linewidth]{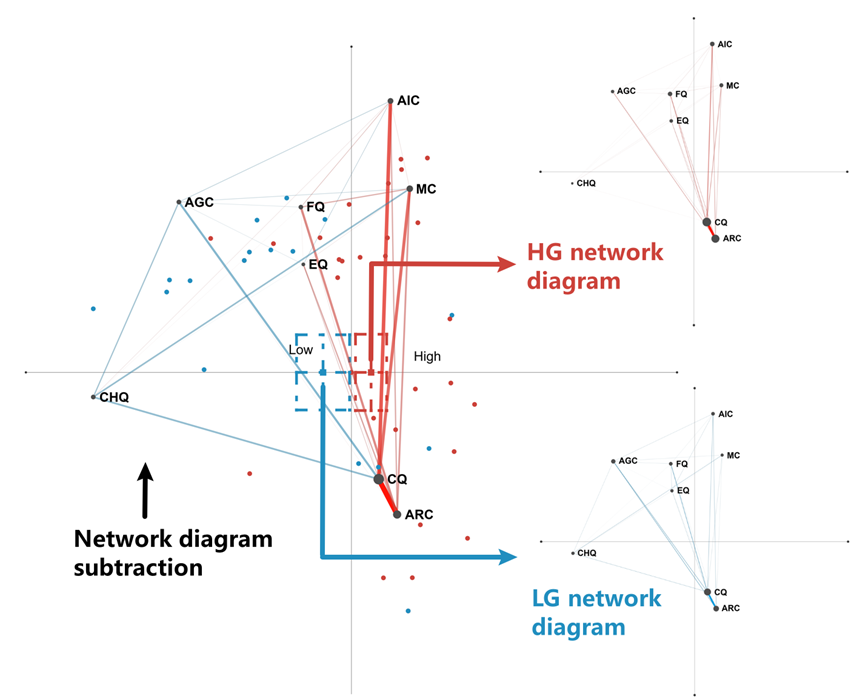}
  \caption{ENA results of LG, HG and network diagram subtraction.}
  \label{fig:ena}
\end{figure}

The High Literacy (HG) network is characterized by strong connections
between AI Improvement Commands (AIC), Clarification Questions (CQ), and
Meta-Commands (MC). This pattern reflects a ``collaborative'' or
``epistemic'' approach. HG students frequently combined requests for text
refinement with questions to clarify underlying concepts (CQ--AIC link).
Furthermore, the integration of Meta-Commands suggests higher
metacognitive regulation, where students actively inquire about optimal
prompting strategies. This indicates an iterative cycle of Sense-Making
$\rightarrow$ Refinement, treating the AI as a co-author and feedback
provider.

\subsection{Statistical Comparison}

To verify these visual differences, we compared the projected ENA scores
(positions in the high-dimensional space). A Mann-Whitney U test
confirmed significant differences between the two groups on the primary
dimension (X-axis) of the ENA space ($U = 683.50$, $p < .001$,
$r = .35$). This statistical evidence confirms that GenAI literacy
levels are associated with fundamentally different behavioral structures
during the learning task.

\section{Discussion and Implications}

\subsection{Behavioral Signatures of GenAI Literacy}

Our findings provide empirical validation for theoretical constructs of
GenAI literacy. The LG group's focus on Generation (AGC) corresponds to
lower-level ``Use \& Apply'' competencies, where the tool is used
primarily to bypass cognitive effort. In contrast, the HG group's focus
on Improvement (AIC) and Clarification (CQ) reflects higher-order
``Evaluate \& Create'' competencies, involving critical refinement and
deep engagement. This suggests that ENA can successfully ``fingerprint''
GenAI literacy through process data, offering a more granular view than
survey-based methods.

\subsection{Implications for Design and Assessment}

These findings have significant implications for the design of
GenAI-integrated learning environments. First, they suggest that literacy
can be assessed \textit{in situ}. Rather than relying on pre-tests,
educational systems could analyze prompt patterns to estimate a learner's
literacy level dynamically.

Second, identifying these patterns enables real-time scaffolding. For
example, systems detecting the ``Generation-Factual'' loop typical of
low literacy could intervene with prompts such as: ``You are focusing on
content generation. Try drafting your own summary first and asking the AI
for critique to improve learning depth.'' Such data-driven interventions
could help shift learners from transactional use to epistemic
collaboration.

\subsection{Conclusion}

This study demonstrates that GenAI literacy is not just a static score
but a dynamic behavioral capability. By applying Epistemic Network
Analysis to interaction logs, we visualized how literacy shapes the
human-AI collaboration process. High literacy facilitates strategic,
refinement-oriented collaboration, while low literacy is characterized by
transactional, generation-oriented dependence. These insights provide a
necessary empirical foundation for developing automated literacy
assessments and adaptive AI tutors.


\section*{Declaration on Generative AI}
  During the preparation of this work, the author used ChatGPT in order to: Grammar and spelling check. After using this tool/service, the author reviewed and edited the content as needed and takes full responsibility for the publication's content.

\bibliography{sample-ceur}

\end{document}